\documentclass{elsart}
\usepackage[dvips]{epsfig}
\usepackage[dvips]{graphicx}
\usepackage{amsfonts}
\usepackage{epsf}
 
\newcommand{\od}[2]{\frac{d#1}{d#2}}
\newcommand{\pd}[2]{\frac{\partial#1}{\partial#2}}

\newcommand{\vol}{\mathop{\mathrm{vol}}}

\newcommand{\area}{\mathop{\mathrm{area}}}

\newcommand{\set}[1]{\left\{ #1 \right\}}

\def\u{\vec{u}}
\def\v{\vec{v}}
\def\V{\bold V}
\def\W{\bold W}

\def\Fl{\bold F}
\def\xl{\bold x}
\def\nl{\bold n}

\def\x{\vec{x}}
\def\b{\vec{b}}

\def\n{\vec{n}}
\def\R{\mathbb{R}}
\def\A{\mathbb{A}}

\def\T{\mathcal{T}}
\def\S{\mathcal{S}}
\def\N{\mathcal{N}}
\def\div{\nabla\cdot}
\def\I{\underline{\underline{I}}}
\def\ssigma{\underline{\underline{\sigma}}}
\def\eps{\underline{\underline{\varepsilon}}}
\def\sgn{\mbox{sgn}}

\begin{document}

\begin{frontmatter}

\title{Tsunami generation by dynamic displacement of sea bed due to dip-slip faulting}

\author{Denys Dutykh, Fr\'ed\'eric Dias}

\address{CMLA, ENS Cachan, CNRS, PRES UniverSud, 61 Av. President Wilson,
F-94230 Cachan, France}

\begin{abstract}
In classical tsunami-generation techniques, one neglects the dynamic sea bed displacement resulting from fracturing of a seismic fault. The present study takes into account these dynamic effects. Earth's crust is assumed to be a Kelvin-Voigt material. The seismic source is assumed to be a dislocation in a viscoelastic medium. The fluid motion is described by the classical nonlinear shallow water equations (NSWE) with time-dependent bathymetry. The viscoelastodynamic equations are solved by a finite-element method and the NSWE by a finite-volume scheme. A comparison between static and dynamic tsunami-generation approaches is performed. The results of the numerical computations show differences between the two approaches and the dynamic effects could explain the complicated shapes of tsunami wave trains.
\end{abstract}


\end{frontmatter}

\section{Introduction}

The accuracy of the computation of the whole life of a tsunami, from generation to inundation, 
obviously depends on the construction of the initial condition. This is why the process of tsunami generation must be 
modelled as accurately as possible. 
Even though the constraint of being able to predict tsunami arrival time, height and location as fast as 
possible must be taken into account (in other words, a trade-off must be found between the precision
and the speed of computation of the initial condition), we believe that so far the scientific community has not
payed enough attention to the crucial subject of tsunami generation.

After the pioneering work of Kajiura \cite{Kajiura1970} it has become a common practice in the tsunami
community to translate the static sea bed
deformation generated by an underwater earthquake onto the free surface and let it propagate. We will refer 
to this method as \textit{passive approach}. The validity of this 
technique was already discussed in \cite{Ohmachi,Dutykh2006}. Three-dimensional (3D) analytical expressions derived from Volterra's formula 
applied to the general study of dislocations \cite{Mansinha,Okada85} are used to
construct the static initial deformation. Similar analytical expressions for two-dimensional (2D) problems were also derived 
by Freund \& Barnett \cite{Freund}, who used the theory of analytic functions of a complex variable. 
The popularity of these analytic solutions can be explained by their relatively
simple explicit form. Thus, their computation is easy and inexpensive. A feature of the solution of
Freund \& Barnett is that nonuniform slip distributions can be considered. In particular, 
slip distributions which remove the singular behavior of the internal stresses at the ends of the slip zone can be dealt with, 
simply by imposing the so-called smooth closure condition on the slip: the slip is zero at the ends.

When simplifying hypotheses such as homogeneity or isotropy are removed, analytical solutions are no longer 
available and the governing equations must be solved numerically. Static deformations caused by slip along a fault 
have been extensively simulated by Masterlark \cite{tim}, who used several dislocation models based on the finite-element 
method (FEM) to estimate the importance of different physical hypotheses. Anisotropy
and heterogeneity turned out to be the most important factors in this
type of modelling. Megna et al. \cite{MBS} also used the FEM to compare numerical results with 
analytical solutions. However neither in \cite{tim} nor in \cite{MBS} were the dynamical aspects and the coupling with
hydrodynamics considered. Moreover the consequences for the resulting tsunami waves were not pointed out.

\begin{figure}[htbp]
\begin{center}
\ifx\JPicScale\undefined\def\JPicScale{1}\fi
\unitlength \JPicScale mm
\begin{picture}(36.25,32.5)(0,0)
\linethickness{0.3mm}
\put(15,15){\line(0,1){17.5}}
\put(15,32.5){\vector(0,1){0.12}}
\linethickness{0.3mm}
\put(15,15){\line(1,0){20}}
\put(35,15){\vector(1,0){0.12}}
\linethickness{0.3mm}
\multiput(15,15)(0.12,0.12){99}{\line(1,0){0.12}}
\put(26.88,26.88){\vector(1,1){0.12}}
\put(36.25,15){\makebox(0,0)[cc]{$y$}}

\put(23.12,26.88){\makebox(0,0)[cc]{$x$}}

\put(12.5,31.25){\makebox(0,0)[cc]{$z$}}

\put(12.5,15.62){\makebox(0,0)[cc]{$O$}}

\linethickness{0.3mm}
\put(0.62,0){\line(1,0){26.88}}
\linethickness{0.3mm}
\multiput(27.5,0)(0.12,0.24){42}{\line(0,1){0.24}}
\linethickness{0.3mm}
\put(32.5,3.75){\line(0,1){6.25}}
\linethickness{0.3mm}
\multiput(27.5,0)(0.16,0.12){31}{\line(1,0){0.16}}
\linethickness{0.3mm}
\multiput(0.62,0)(0.12,0.24){42}{\line(0,1){0.24}}
\linethickness{0.3mm}
\put(5.62,10){\line(1,0){26.88}}
\put(-3.12,-1.88){\makebox(0,0)[cc]{$-\frac{L}{2}$}}

\put(28.12,-3.12){\makebox(0,0)[cc]{$\frac{L}{2}$}}

\put(31.25,4.38){\makebox(0,0)[cc]{$\delta$}}

\linethickness{0.3mm}
\multiput(15,0)(0,2){8}{\line(0,1){1}}
\put(13.75,-1.88){\makebox(0,0)[cc]{$-d$}}

\put(0,6.25){\makebox(0,0)[cc]{$W$}}

\put(5,3.75){\makebox(0,0)[cc]{}}

\end{picture}
\end{center}
  \caption{Geometry of the source model. The fault has width $W$, length $L$, depth $d$ and dip angle $\delta$.}\label{fig:sketchokada}
\end{figure}
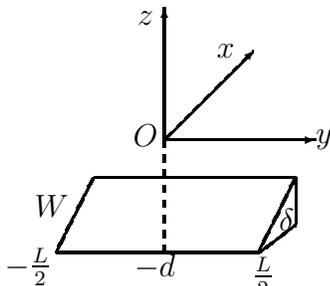

When one uses as initial condition a static seismic source together with 
the translation of the sea bed deformation onto the free surface, one
neglects the rupture velocity and the rise time. Several studies have already been performed
to understand wave formation due to different prescribed
bottom motions by introducing either some type of rise time or some type of rupture velocity.
For example, Todorovska \& Trifunac \cite{todo} studied the generation of waves by a slowly spreading uplift of the bottom.
Hammack \cite{Hammack} generated waves experimentally by raising or lowering a box at one end of a channel,
and considered various laws for the rise or the fall of the box. In their review paper, Dutykh \& Dias \cite{Dutykh2007}
generated waves theoretically by multiplying the static deformations caused by slip along a fault by various time laws:
instantaneous, exponential, trigonometric and linear. Haskell \cite{Haskell} was one of the first to take into account the
rupture velocity. In fact he considered both rise time, $T$, and rupture velocity, $V$. Consider the source shown in Figure
\ref{fig:sketchokada}. The two horizontal coordinates $x$ and $y$, and the vertical coordinate $z$ are denoted by
$\x=(x,y,z)$. 
Let $\b(\x,t)$ denote the fault displacement
function and $\b_0(\x)$ the final displacement. The following form for $\b(\x,t)$ was considered by Haskell:
\begin{equation} \label{Haskell_b}
\b(\x,t) = \left\{ \begin{array}{cl} 0 & t-\zeta/V < 0 \\ (\b_0/T) (t-\zeta/V) & 0 < t-\zeta/V < T \\ \b_0 & t-\zeta/V > T 
\end{array} \right.
\end{equation}
The coordinate $\zeta$ is a coordinate along the fault.
Eq. (\ref{Haskell_b}) implies that at $t=0$ a fracture front is established instantaneously over a width $W$ of the $y-$axis
at depth $d$. The front propagates
unilaterally at constant velocity $V$ over a length $L\cos\delta$ of the $x-$axis. At any fixed point on the fault plane the relative
displacement increases at constant velocity from $0$ at $t=\zeta/V$ to a constant final value $\b_0$ at $t=T+\zeta/V$. 
Okumura \& Kawata \cite{Okumura} used Haskell's approach
to investigate the effects of rise time and rupture velocity on tsunami generation. They considered two cases of sea 
bottom motion: (i) with only
rise time and (ii) with both rise time and rupture velocity. They found that the effects of rupture
velocity are much smaller than those of rise time when the rise time is assumed to be long (over 10 min). Ohmachi et al. 
\cite{Ohmachi} also considered rise time and rupture velocity but unfortunately the dynamics is not clearly explained
in their paper. Apparently they did not solve the elastodynamic equations with the second-order time derivative (see next section).
Another attempt to understand dynamical effects is that of Madariaga \cite{Madariaga}, who considers a dip-slip
dislocation propagating in a half-space and solves the elastodynamic equations by using the double Laplace transform. The
solution is elegant but rather complex. Unfortunately no plots of the deformation of the 
free surface are provided and the coupling with the water layer is not considered either. 
The present study can be considered as an attempt 
to understand the coupling between seismic faulting and hydrodynamics by integrating numerically
the time-dependent elasticity equations as well as the time-dependent fluid equations.
The authors have already adressed the problem of tsunami generation in 
\cite{Dutykh2006,Dutykh2007}. The main feature of the present 
study is the use of a more realistic earthquake source model. 

The paper is organised as follows. In Section \ref{section:mathmodel} 
we briefly describe the mathematical models, both for solid and fluid motions.  
Section \ref{section:numerical} contains details on the numerical 
methods used to solve the governing equations. The numerical
method for the solid motion is validated in Section \ref{section:validation}. Section \ref{section:numeric} provides a comparison between the traditional approach to
tsunami generation (in which the static sea bed deformation is translated
onto the free surface) and the more realistic approach of dynamic generation (in which the wavetrain is generated
by the motion of the bottom). We reveal numerically that the dynamic generation of tsunamis can for example create
a leading depression wave when one expects a leading elevation wave.


\section{Mathematical models}\label{section:mathmodel}

Even though the numerical results shown in this paper are for 2D configurations, the modeling is performed for 
3D problems. The horizontal coordinates are denoted by $x$ and $y$, while the vertical coordinate is denoted
by $z$. The displacements are denoted by $u_x, u_y$ and $u_z$. We use different origins along the vertical axis for the
solid and fluid motions. In the earth domain, $z=0$ denotes the sea bed at rest (assumed to be flat) while in the fluid domain, $z=0$ denotes the sea surface at rest.

\subsection{Dynamic fault model}

The fault is assumed to be inside a geological viscous medium.
Earth's crust is assumed to be a viscoelastic material of density $\rho$. We choose 
the Kelvin-Voigt viscosity model \cite{Quiblier1997} which consists 
in using complex elastic coefficients (with negative imaginary parts in order to dissipate
wave energy). For isotropic media it means that the Lam\'{e} coefficients have a nonpositive imaginary part:
$\lambda^* = \lambda_r - i\lambda_i, \quad \mu^* = \mu_r - i\mu_i, $
where $\lambda_r, \mu_r > 0$ and $\lambda_i, \mu_i \geq 0$. 
The classical elasticity equations are obtained by choosing $\lambda_i \equiv 0$ and $\mu_i \equiv 0$.
On the time-scales relevant to our problem, elasticity is sufficient and the assumption of a 
Kelvin-Voigt viscous material is unnecessary. But we keep it for the sake of completeness.

Let $c_P$ and $c_S$ be the classical velocities for the propagation 
of $P$ and $S$ waves in a medium of density $\rho$:
$$
  c_P = \sqrt{\frac{\lambda_r+2\mu_r}{\rho}}, \quad
  c_S = \sqrt{\frac{\mu_r}{\rho}}.
$$
Complex Lam\'{e} coefficients yield complex velocities for wave propagation,
$$
  c_P^* = c_P\sqrt{1+{i}/{Q_P}}, \quad 
  c_S^* = c_S\sqrt{1+{i}/{Q_S}},
$$
where the coefficients $Q_P$ and $Q_S$ are defined as follows:
$$
  Q_P = -\frac{\lambda_r+2\mu_r}{\lambda_i+2\mu_i}, \quad
  Q_S = -\frac{\mu_r}{\mu_i}.
$$
The factors $Q_P$ and $Q_S$ measure the viscosity of the geological medium. In this study we restrict
our attention to the weakly viscous case (${1}/|{Q_P}| \ll 1$ and ${1}/|{Q_S}| \ll 1$).

Let $\ssigma$ represent the stress tensor. The displacement field $\u(x,y,z,t)=(u_x,u_y,u_z)$ satisfies the classical
elastodynamic equations from continuum mechanics:
\begin{equation}
  \div\ssigma = \rho\pd{^2\u}{t^2}.
\end{equation}
It is common in seismology to assume that the stress tensor $\ssigma$ 
is determined by Hooke's law through the strain tensor 
$\eps = \frac12\Bigl(\nabla\u + \nabla^t\u\Bigr)$. Therefore
\begin{equation}
  \ssigma = \lambda^*(\div\u)\I + 2\mu^*\eps.
\end{equation}
Thus, we come to the following linear viscoelastodynamic problem\footnote{%
We use the prefix ``visco-'' due to the presence of the imaginary part 
in the Lam\'{e} coefficients, which is responsible for small wave damping.}:
\begin{equation}\label{eq:dynelast}
  \div\left(\lambda^*(\div\u)\I + \mu^*(\nabla\u+\nabla^t\u)\right) = \rho\pd{^2\u}{t^2}.
\end{equation}

Recall that the mechanical characteristics $\rho$, $\lambda^*$ and
$\mu^*$ can possibly depend on the spatial coordinates $(x,y,z)$. 
However we will assume that they are constant in the numerical applications.

The fault is modeled as a dislocation inside a viscoelastic material.
This type of model is widely used for the interpretation of 
seismic motion. A dislocation is considered as a surface (in 3D problems)
or a line (in 2D problems) in a continuous medium where the displacement field
is discontinuous. The displacement vector is increased by the amount of the
Burgers vector $\b$ along any contour $C$ enclosing the dislocation surface (or line), i.e.
\begin{equation}
  \oint\limits_{C} d\u = \b.
\end{equation}
We let a dislocation run at speed $V$ along a fault inclined at an angle $\delta$ with respect to the
horizontal. Rupture starts at position $x=0$ and $z=-d$ (it is supposed to be infinitely long in the transverse
$y-$direction) and propagates with constant rupture speed $V$ for a finite time $L/V$ in the direction $\delta$
stopping at a distance $L$. Let $\zeta$ be a coordinate along the dislocation line. On the fault located in the 
interval $0 < \zeta < L$ slip is assumed to be constant. The rise time is assumed to be $0$.

\subsection{Fluid layer model}

Since the main purpose is to model tsunami generation processes and since tsunamis are 
long waves, it is natural to choose the nonlinear shallow
water equations (NSWE) as hydrodynamic model. These equations are widely used
in tsunami modelling, especially in codes for operational use \cite{TS,Syno2006}. The validity of the NSWE model and
the question of the importance of dispersive effects have already been addressed by the authors in \cite{Kervella2007}.

Let $\eta$ denote the free-surface elevation with respect to the still
water level $z=0$, $\v=(v_x,v_y)$ the horizontal velocity vector, $g$ the acceleration due to gravity and $z=-h(x,y,t)$ the
bathymetry. The NSWE in dimensional form read
\begin{equation} \label{stvenant}
  \pd{\eta}{t} + \div\left((h+\eta)\v\right) = -\pd{h}{t}, \qquad
  \pd{\v}{t} + \frac12\nabla|\v|^2 + g\nabla\eta = 0.
\end{equation}
The effect of the moving bottom appears in the source term $-\partial h/\partial t$ in the first equation.
The unknowns $\eta$ and $\v$ are
functions of time and of the horizontal coordinates $x$ and $y$. Since the NSWE are essentially obtained 
from depth-averaging the Euler equations, the dependence on the vertical coordinate $z$ disappears from the equations.
The coupling between the earth and fluid models is made through
the function $h(x,y,t)$ which describes the moving sea bottom bathymetry.


\section{Numerical methods}\label{section:numerical}

In the present study we made two natural choices. The solid
mechanics equations are solved using the FEM with fully implicit time integration,
while for the hydrodynamic part we take advantage of the hyperbolic structure
of the governing equations and use a solver based on the finite-volume
scheme (see for example \cite{Benkhaldoun2006,Kervella2007}).

\subsection{Discretization of the viscoelastodynamic equations}

In order to apply the FEM one first rewrites the governing equation
(\ref{eq:dynelast}) in variational form. The time-derivative operator is discretized through a classical 
second-order finite-difference scheme. The method we use is fully implicit and has the advantage of being free
of any Courant-Friedrichs-Lewy-type condition. In such problems implicit schemes become 
advantageous since the velocity of propagation of seismic waves
is of the order of 3 to 4 km/s. We apply the $\mathbb{P}2$ finite-element discretization procedure.
For the numerical computations, the freely available code FreeFem++ \cite{Hecht1998} is used.

Let us say a few words about the boundary conditions and the treatment
of the dislocation in the program. Concerning the boundary conditions, 
we assume that the sea bed is a free surface, that is 
$\ssigma\cdot\n = \vec{0}$ at $z = 0$. 
The other boundaries
are assumed to be fixed or, in other words, Dirichlet type boundary conditions 
$\u = \vec{0}$ are applied. The authors are aware of the reflective properties of this type
of boundary conditions. In order to avoid the reflection of seismic
waves along the boundaries during the simulation time, 
we take a computational domain which is sufficiently large. This approach is not computationally
expensive since we use adaptive mesh algorithms \cite{Hecht1998} and in the regions 
far away from the fault, element sizes are considerably bigger than in the fault vicinity. 

Next we discuss the implementation of the dislocation surface. Across the fault, 
the displacement field is discontinuous and satisfies the following relation:
\begin{equation}\label{eq:condburgers}
  \u^+(\x,t) - \u^-(\x,t) = \b(\x,t),
\end{equation}
where the signs $\pm$ denote the upper and lower boundary of the dislocation surface,
respectively. The propagation of Burger's vector along the fault is given by
\begin{equation} \label{rupt_model}
 \b(\x,t)= \b_0 H(t-\zeta/V), 
\end{equation}
where $V$ is the rupture velocity, $H$ the Heaviside unit step function and $\zeta$ a coordinate along the dislocation line.  


\subsection{Finite-volume scheme for the nonlinear shallow-water equations}

Here we briefly describe the discretization of the model (\ref{stvenant}) by a standard cell-centered finite volume method. 
The system (\ref{stvenant}) can be written as
\begin{equation} \label{stvenant2}
  \pd{\V}{t} + \nabla\cdot \Fl(\V) = \S \,,
\end{equation}
where $\V = (\eta, u, v)$ is the vector of conservative variables (which coincide here with the physical variables), $\S = (-\partial h/\partial t, 0, 0)$ the source term, and, for every $\nl \in \R^2$, 
\begin{equation} \label{Fn}
  \Fl(\V)\cdot\nl = \left( (h+\eta) (\v\cdot\nl), \left(\frac12 |\v|^2 + g\eta\right) \nl \right).
\end{equation}
The Jacobian matrix $\A(\V) \cdot \nl$ is defined by
\begin{equation} \label{mat_jacob}
\A(\V) \cdot \nl = \pd{\bigl(\Fl(\V)\cdot\nl\bigr)}{\V} = \left( \begin{array}{cc}
\v\cdot\nl & (h+\eta)\nl \\
            g\nl & \v\otimes\nl
       \end{array} \right).
\end{equation}

The computational domain $\Omega\subset\R^2$ is triangulated into a set of control volumes $\Omega=\cup_{K\in\T} K$.
We integrate Eq. (\ref{stvenant2}) on $K$:
\begin{equation}\label{eq:conservlaw}
	\od{}{t}\int_{K} \V \;d\Omega + \sum_{L\in\N(K)}\int_{K\cap L}\Fl(\V)\cdot\nl_{KL} \;d\sigma
  	  = \int_{K} \S \;d\Omega\,,
\end{equation}
where $\nl_{KL}$ denotes the unit normal vector on $K\cap L$ pointing into $L$ and $
  \N(K) = \set{L\in\T: \area(K\cap L) \neq 0}\,.
$
Then, setting
$$
 \V_K(t) := \frac{1}{\vol(K)}\int_{K} \V(\xl,t) \;d\Omega \;,
$$
we approximate (\ref{eq:conservlaw}) by
\begin{equation}\label{a_resoudre}
	\od{\V_K}{t} + \sum_{L\in\N(K)} \frac{\area(L\cap K)}{\vol(K)} \Phi(\V_K, \V_L; \nl_{KL}) =  \S_K\;,
\end{equation}
where the numerical flux $$\Phi(\V_K, \V_L; \nl_{KL}) 
\approx\frac{1}{\area(L\cap K)}\int_{K\cap L}\Fl(\V)\cdot\nl_{KL} \;d\sigma$$ 
is explicitly computed by the FVCF formula of Ghidaglia {\it et al.} \cite{Ghidaglia2001}:
\begin{equation}\label{CFFV}
\Phi(\V, \W; n)=\frac{\Fl(\V) \cdot\nl+\Fl(\W) \cdot\nl}{2}-\sgn(\A_n(\mu(\V,\W)))\frac{\Fl(\W) \cdot\nl-\Fl(\V) \cdot\nl}{2}\,.
\end{equation}
Here the Jacobian matrix $\A_n(\mu)$ is defined in (\ref{mat_jacob}), $\mu(\V,\W)$ is an arbitrary mean between $\V$ and $\W$ and $\sgn(M)$ is the matrix whose eigenvectors are those of $M$ but whose eigenvalues are the signs of that of $M$.

In our problem, the Jacobian matrix (\ref{mat_jacob}) has three distinct eigenvalues
$$ \lambda_{1,3} = \v\cdot\nl \pm c_s, \quad \lambda_2 = 0,
$$
where $c_s = \sqrt{g(h+\eta)}$ is the velocity of long gravity waves.
The right and left eigenvectors can be easily computed. Then it is straightforward to compute the sign matrix
appearing in (\ref{CFFV}) and the numerical scheme is thus completely defined.

In this section we did not deal with boundary 
conditions. This is a complicated topic in finite-volume methods (see \cite{Ghidaglia2005} for details).

\section{Validation of the numerical method}\label{section:validation}

In this section we consider an analytic solution to the line dislocation
problem in the static case. Use is made of the
well-known result described for example by Freund \& Barnett \cite{Freund} or Okada \cite{Okada85}. 
In order to simplify the expressions, we only consider the 2D case (in other words, the fault
is infinite in the $y-$direction).
In fact the best expression is that given by equation (24) in \cite{Madariaga}. We checked
that it is in full agreement with the limit of Okada's solution as the width becomes infinite.
The sketch of the domain is given in Figure \ref{fig:sketchokada}. The fault has infinite width ($W \to \infty$). 
Its length is $L$, its depth $d$ and its dip angle $\delta$.

In the present paper we only give
the vertical displacement component $u_z$ along the free surface, since it plays the 
most important role in tsunami formation. It can be expressed as the difference between two 
contributions, that from a first dislocation located at the beginning of the fault and that from a second
dislocation located at the end of the fault. Let $d_L=d-L\sin\delta$. One has
\begin{equation}
u_z = |\b_0| \left[ U_z\left(\frac{x}{d},\delta\right) - U_z\left(\frac{x-L\cos\delta}{d_L},\delta\right) \right], 
\end{equation}
where
\begin{equation}\label{eq:solution}
U_z\left(\frac{x}{d},\delta\right) = \frac{1}{\pi} \left[ \sin\delta \arctan\frac{x}{d} - 
\frac{d(d\cos\delta-x\sin\delta)}{x^2+d^2} \right].
\end{equation}

\begin{figure}
	\centering
		\includegraphics[width=0.60\textwidth]{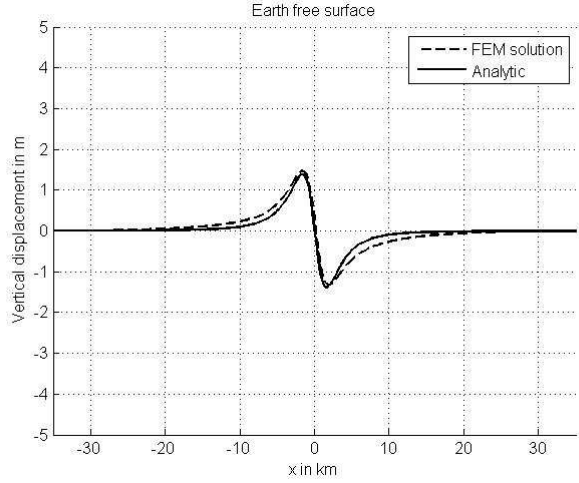}
	\caption{Comparison between analytical and numerical solutions for a static 2D fault
with a dip angle equal to $\pi/2$.}
	\label{fig:comparison}
\end{figure}

For the validation of our numerical method we chose a fault
corresponding to a dip angle $\delta={\pi}/{2}$. The values of the
other parameters are given in Table~\ref{tab:params}. This problem
was solved by FEM after neglecting the dynamic terms. The results
of the comparison with solution (\ref{eq:solution}) are given in Figure
\ref{fig:comparison}. Good qualitative and quantitative agreement can be
seen. Megna et al. \cite{MBS}, who also considered static displacement due to uniform slip
across a normal fault, compared the 2D FEM results with the analytical solution
in the case of a normal fault. In their conclusion, they state that it is for the vertical
component of the surface displacement that the discrepancies are the largest.

\section{Results of the simulation}\label{section:numeric}

\begin{figure}
	\centering
		\includegraphics[width=0.8\textwidth]{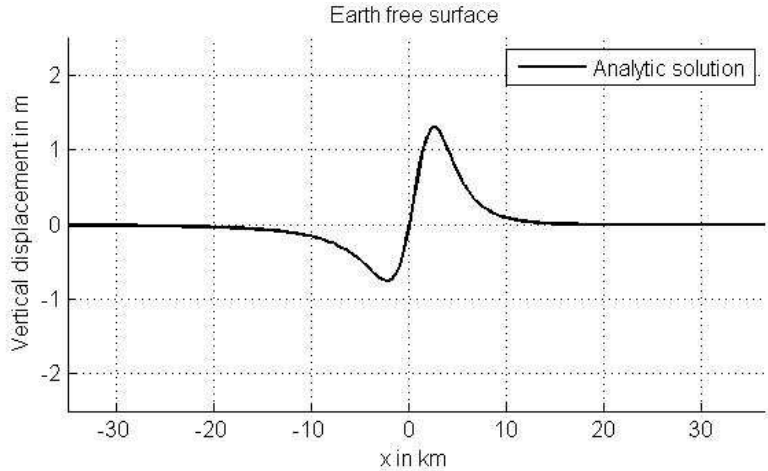}
	\caption{Static deformation due to the dislocation corresponding to the parameters given in Table \ref{tab:params}.}
	\label{fig:static}
\end{figure}

In this section, we use the set of physical parameters given in 
Table \ref{tab:params}. The static sea bed deformation obtained with the analytical solution is depicted in Figure \ref{fig:static}. 
Note that the only difference between Figures \ref{fig:comparison} and \ref{fig:static} is the value of the dip angle. 


\begin{table}
\begin{center}
\begin{tabular}{lc}
\hline
\bf{Parameter} & \bf{Value} \\
\hline
  Young modulus, $E$, GPa & 9.5 \\
  Poisson ratio, $\nu$ & 0.27 \\
  Damping coefficient, $\lambda_i$ & 500 \\
  Damping coefficient, $\mu_i$ & 200 \\
  Fault depth, $d$, km & 4 \\
  Fault length, $L$, km & 2 \\
  Dip angle, $\delta$, $^\circ$ & 13 \\
  Burger's vector length, $|\b_0|$, m & 10 \\
  Water depth (uniform), $h_0$, m & 400 \\
  Acceleration due to gravity, $g$, m/s$^2$ & 9.8 \\
\hline
\end{tabular}
\caption{Parameters used in the numerical computations. The water depth and 
the spatial extent in the main direction of propagation were chosen so that dispersive effects can be neglected.}
\label{tab:params}
\end{center}
\end{table}
In order to illustrate the numerical computations we chose several test cases of active/passive tsunami generation.
The passive generation approach consists in translating the
static sea bed deformation onto the free surface and letting it propagate under gravity \cite{Kajiura1970}.
On the other hand, the active approach uses the bottom motion for wave generation. We proceed 
by computing the first eight or fifteen seconds of the earthquake dynamics. 
Then the bottom configuration is assumed to remain frozen during the rest of the simulation.
Concerning the dynamical aspects of rupture propagation, we consider the Heaviside-type approach (\ref{rupt_model}) 
where the dislocation propagates along the
fault with rupture velocity $V$. One could use instead a dislocation for which Burger's vector $\b_0$ is also space-dependent.
But the main goal of the present study is to make an attempt to include the dynamic displacement of the sea bed. In
the dynamical approach, we consider three cases: the limiting case where the rupture velocity $V$ is infinite, a fast event
with $V=2.5$ km/s and a slower event with $V=1$ km/s.  

We show below the differences
between the passive and the dynamic approaches. This question has already been addressed by the authors \cite{Dutykh2006}
in the framework of the linearized potential flow equations and of a simplified model for bottom deformation. 

\begin{figure}
	\centering  
 \includegraphics[width=0.60\textwidth]{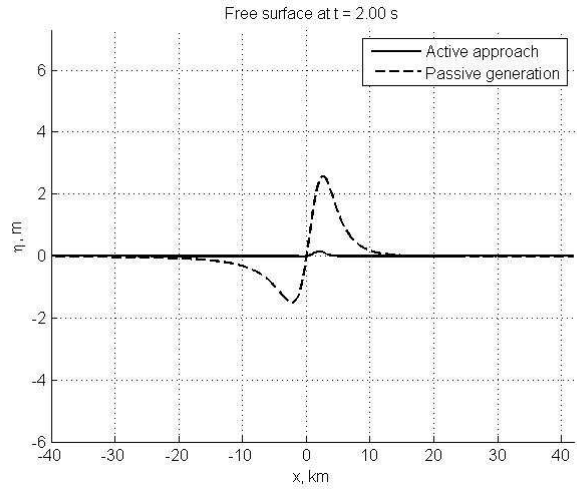}
	\caption{Water free surface at the beginning of the earthquake ($t=2$s) according to two approaches
	of tsunami generation: passive versus active (with infinite rupture velocity).}
	\label{fig:t0-75}
\end{figure}

\begin{figure}
	\centering 
 \epsfig{file=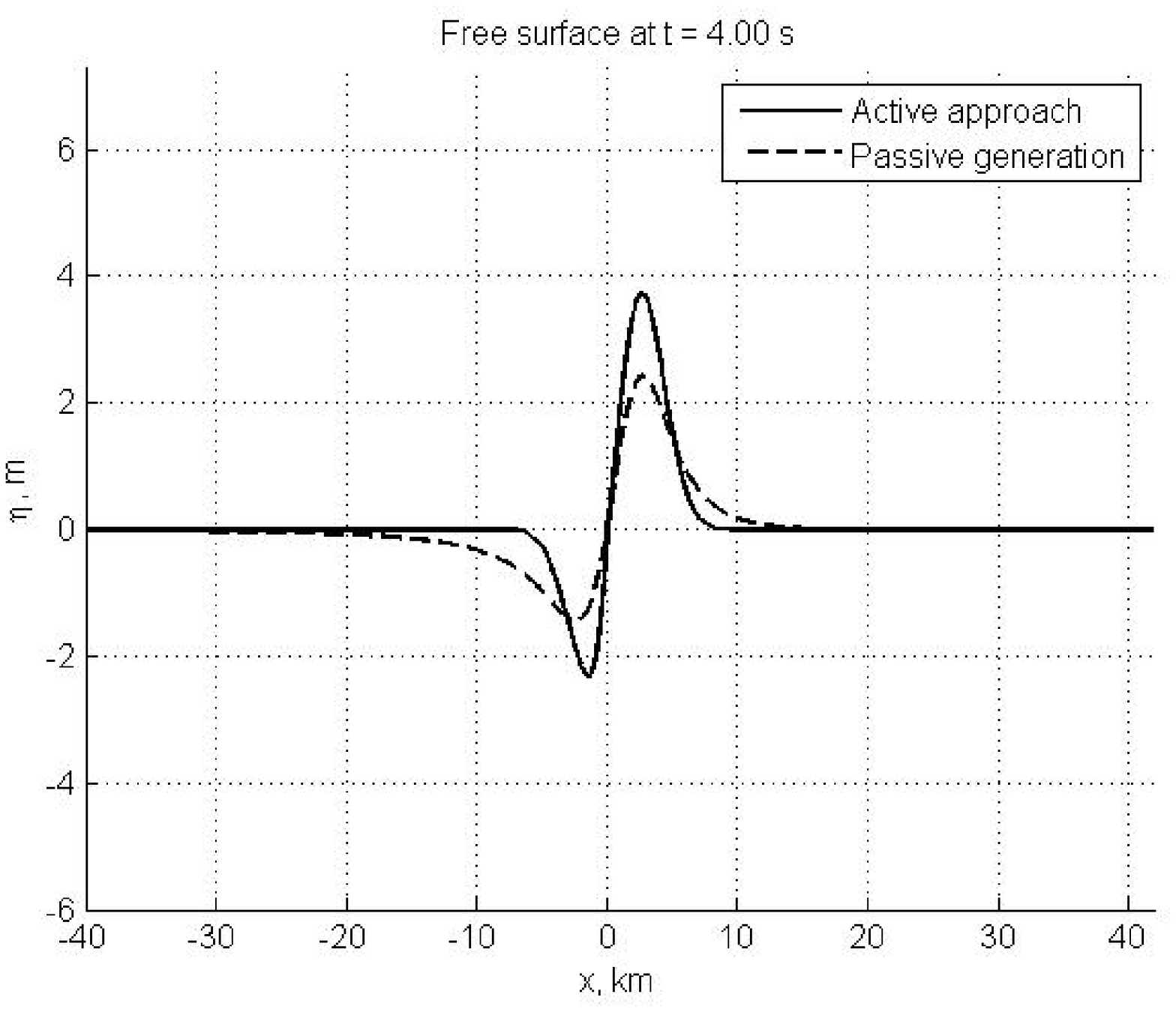, width=7cm} \epsfig{file=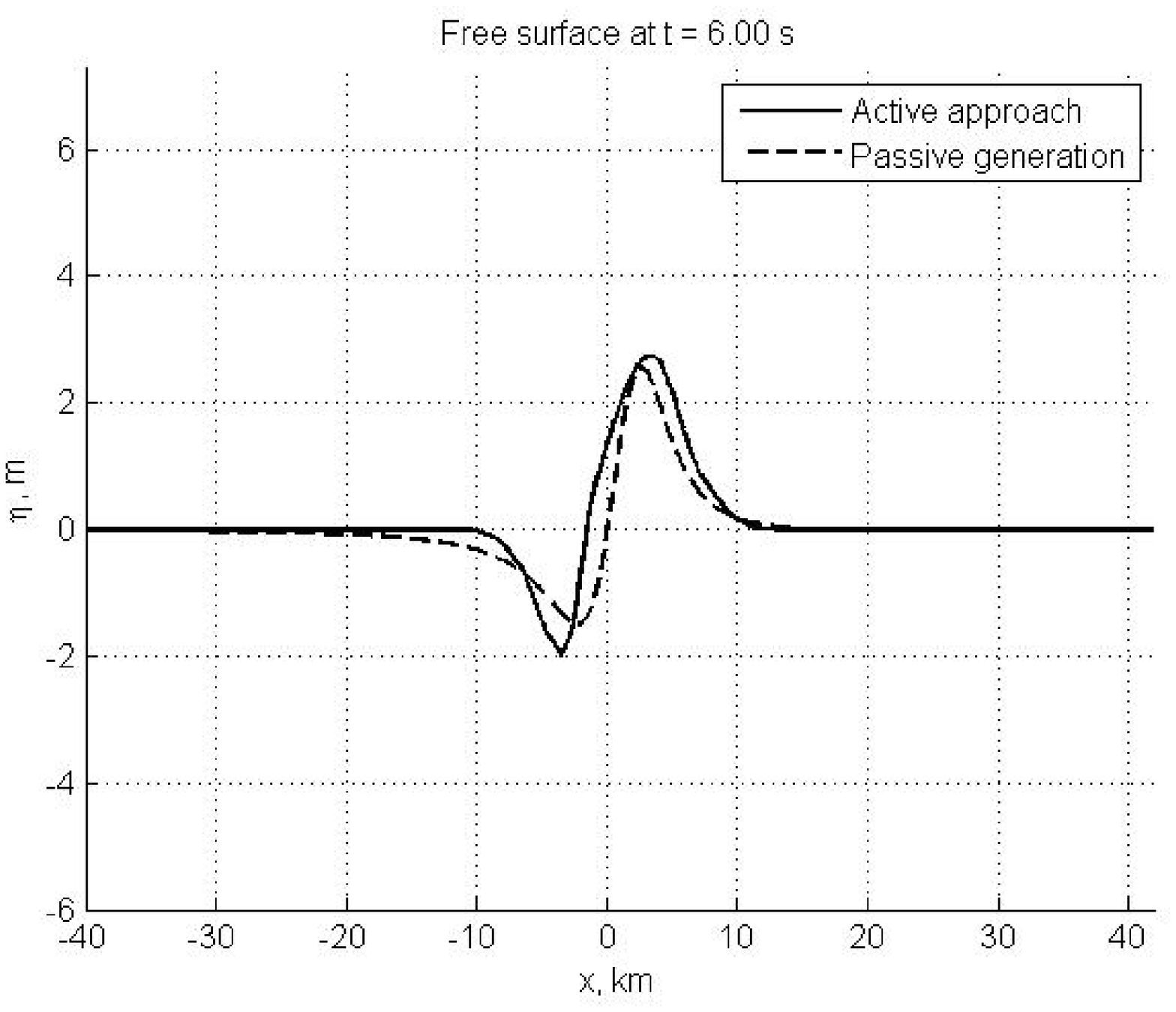, width=7cm}
	\caption{Same as figure \ref{fig:t0-75} for times $t=4$s and $t=6$s.}
	\label{fig:t2-0}
\end{figure}

\begin{figure}
	\centering 
 \epsfig{file=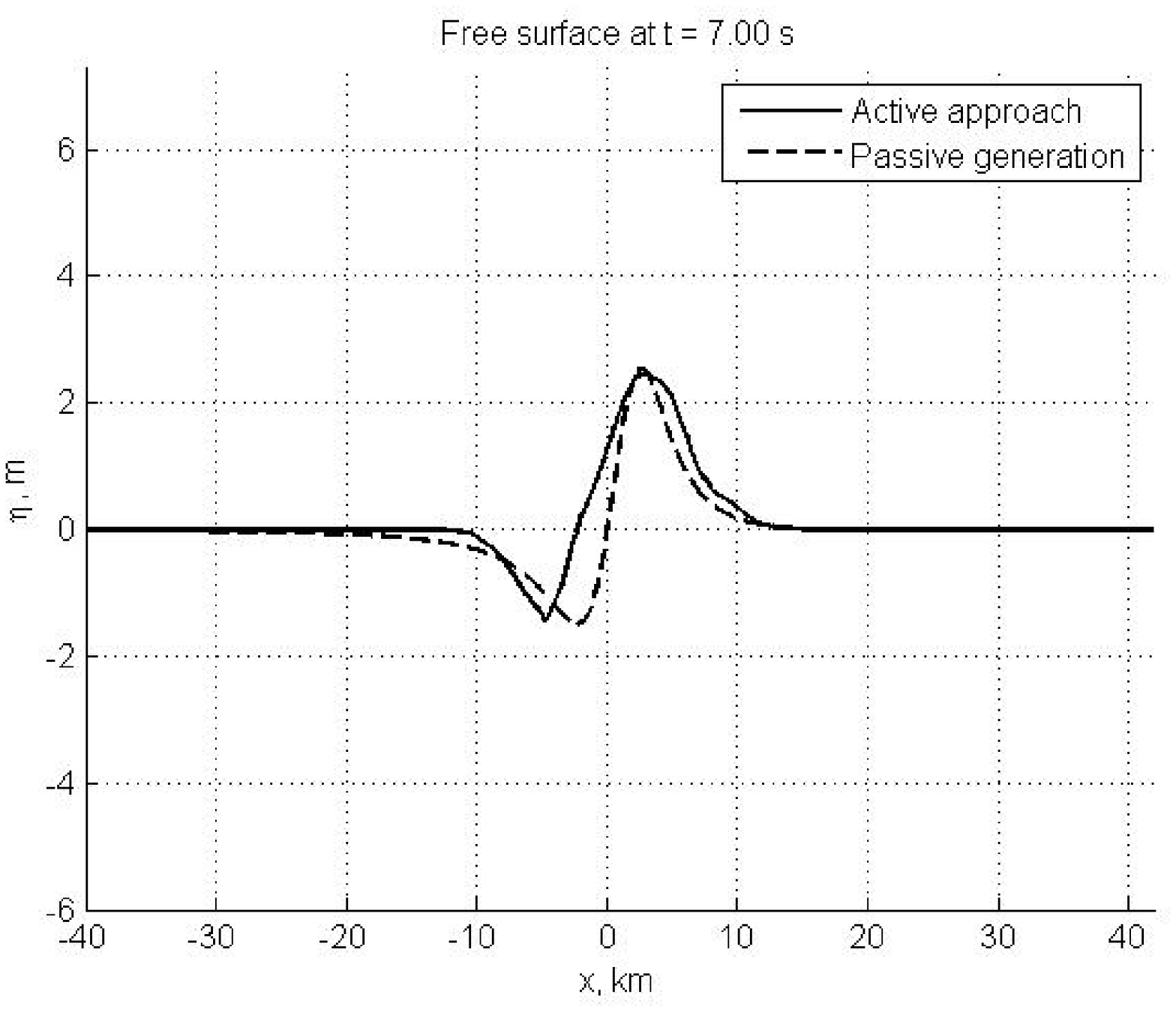, width=7cm} \epsfig{file=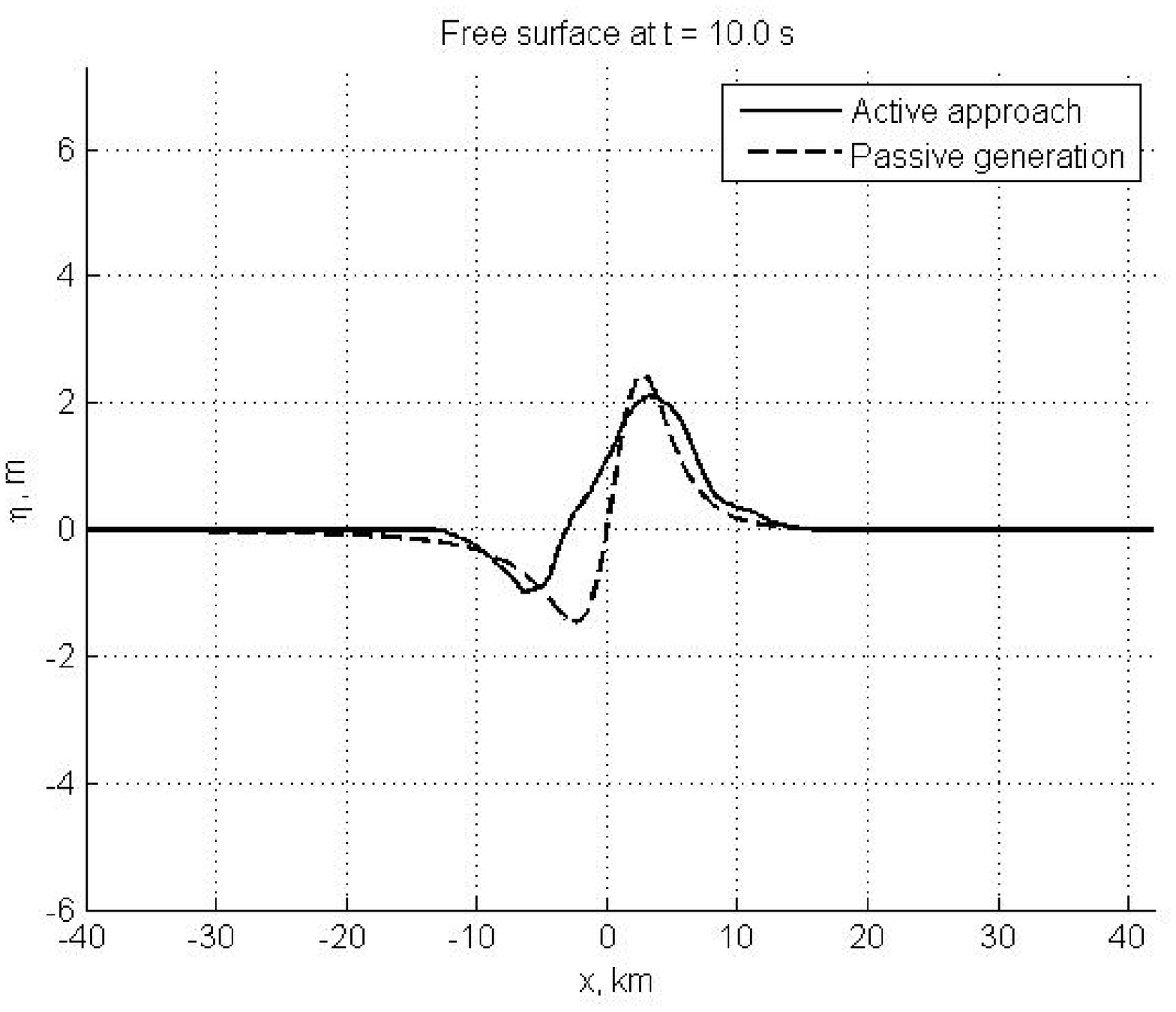, width=7cm}
	\caption{Same as figure \ref{fig:t0-75} for times $t=7$s and $t=10$s.}
	\label{fig:t4-0}
\end{figure}

In the first comparison
we use a strong coupling between the dynamic displacement of the sea bed and the fluid layer equations and compare it with
the passive approach, in which the static solution shown in Figure \ref{fig:static} is translated onto the
free surface as initial condition. The rupture velocity $V$ is assumed to be infinite. Moreover
the earthquake dynamics is computed during the first eight seconds. 
The free surface at the beginning of the tsunami generation process is shown 
on Figure \ref{fig:t0-75}. Further steps of this process are given in Figures
\ref{fig:t2-0}-\ref{fig:t4-0}. One may have the impression that the passive solution
does not evolve. In fact, the explanation lies in the presence of two different time
scales in this problem. The fast time scale is provided by the earthquake ($P-$ and $S-$waves) 
and the slow one by water gravity waves. Since the active generation solution is directly
coupled to the bottom dynamics, it evolves with the fast time scale.
It is interesting to compare Figures \ref{fig:t2-0} and \ref{fig:t4-0}. 
One can see that the active approach gives at the beginning an amplitude which is almost twice larger 
but the amplitudes become comparable a few seconds later.

\begin{figure}
	\centering 
 \epsfig{file=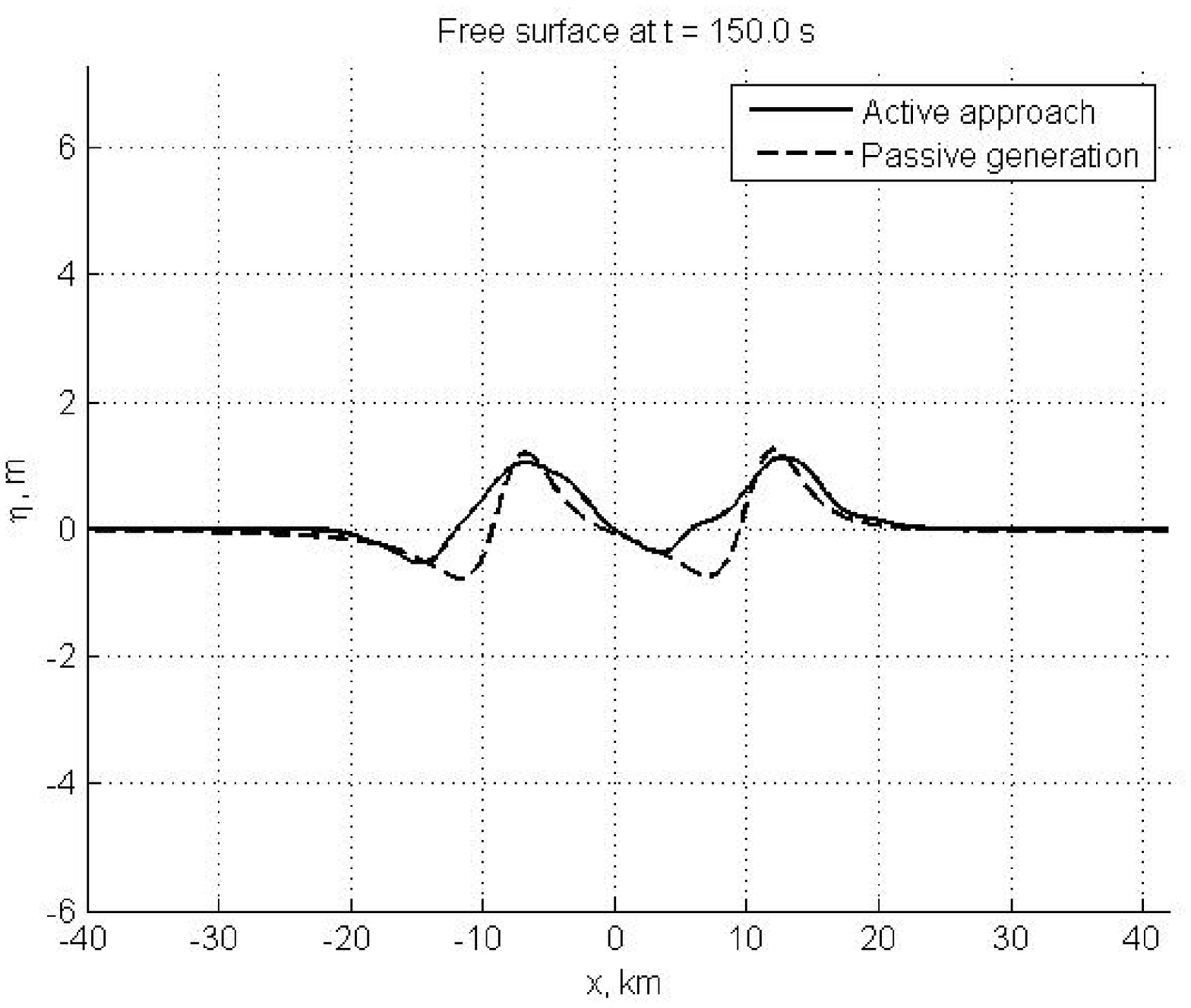, width=7cm} \epsfig{file=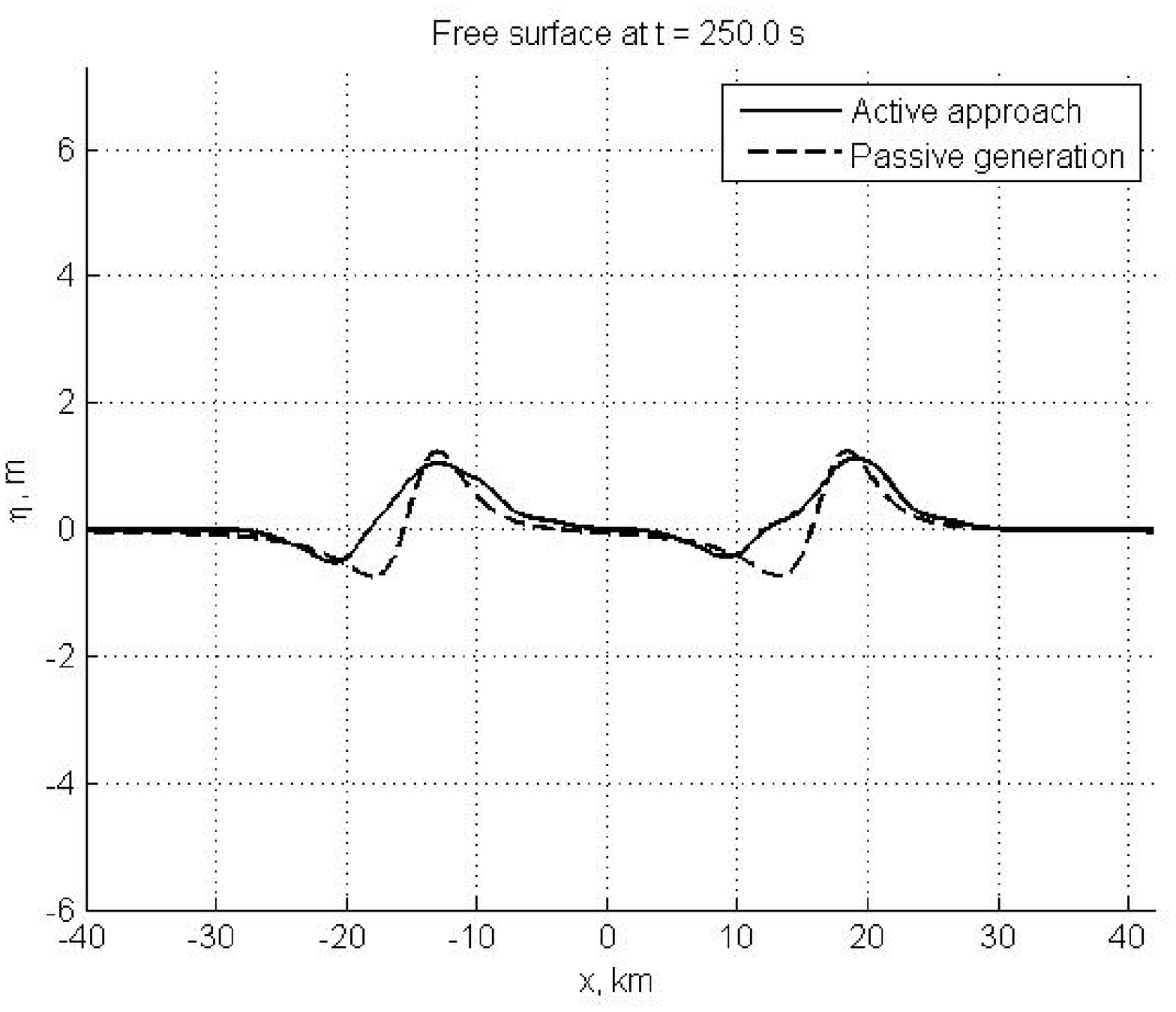, width=7cm}
	\caption{Same as figure \ref{fig:t0-75} for times $t=150$s and $t=250$s. The wave 
	is leaving the generation zone (left plot) and starting to propagate (right plot).}
	\label{fig:t20-1}
\end{figure}

The free-surface elevations are computed until the wave enters the purely propagation
stage. This corresponds to Figure \ref{fig:t20-1}. One notices
that the resulting wave amplitude and velocity are almost the same. Of course 
the waveform is different. One can see as well that the location of the elevation
wave is the same, while the depression wave is slightly shifted. It can be explained
by the larger extent of the dynamic solution.
Thus, we can conclude from this first comparison that if one is only interested in tsunami travel 
time or even in rough inundation zone estimation, the passive approach can be used. 

\begin{figure}
	\centering 
 \epsfig{file=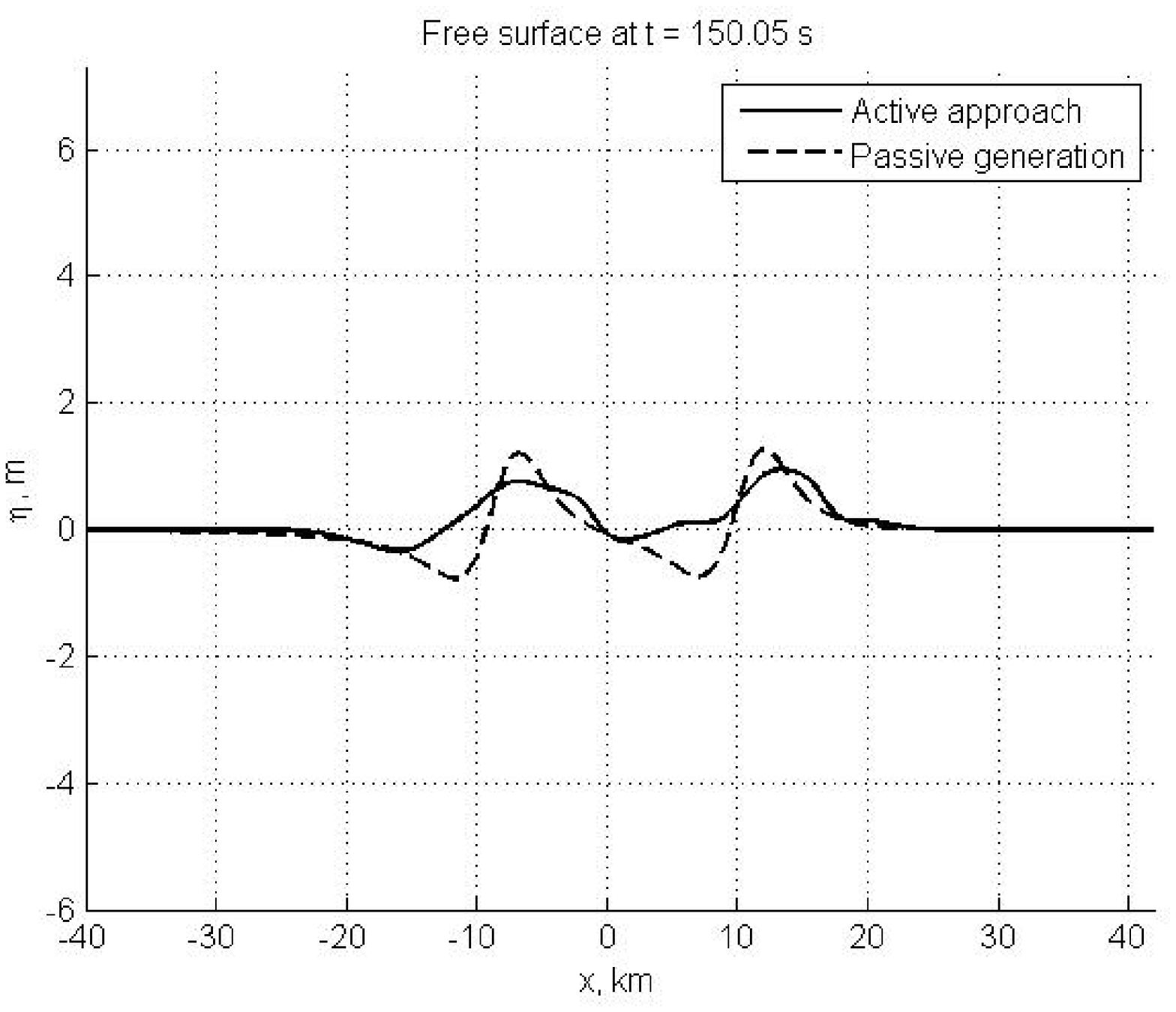, width=7cm} \epsfig{file=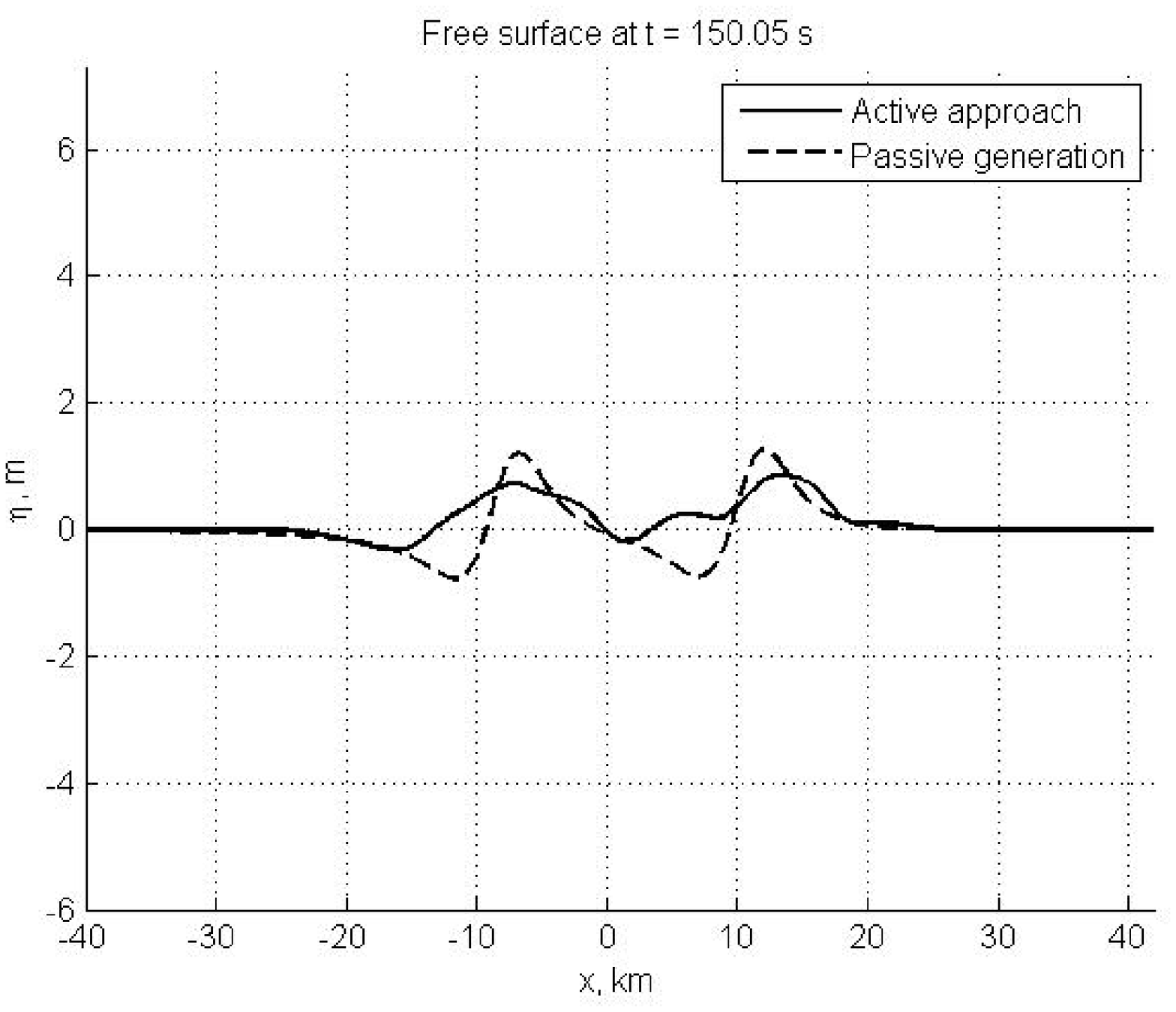, width=7cm}
	\caption{Same as figure \ref{fig:t20-1} (left plot) for two rupture velocities: $V=1$ km/s (left)
and $V=2.5$ km/s (right).}
	\label{fig:sf150}
\end{figure}

\begin{figure}
	\centering 
 \epsfig{file=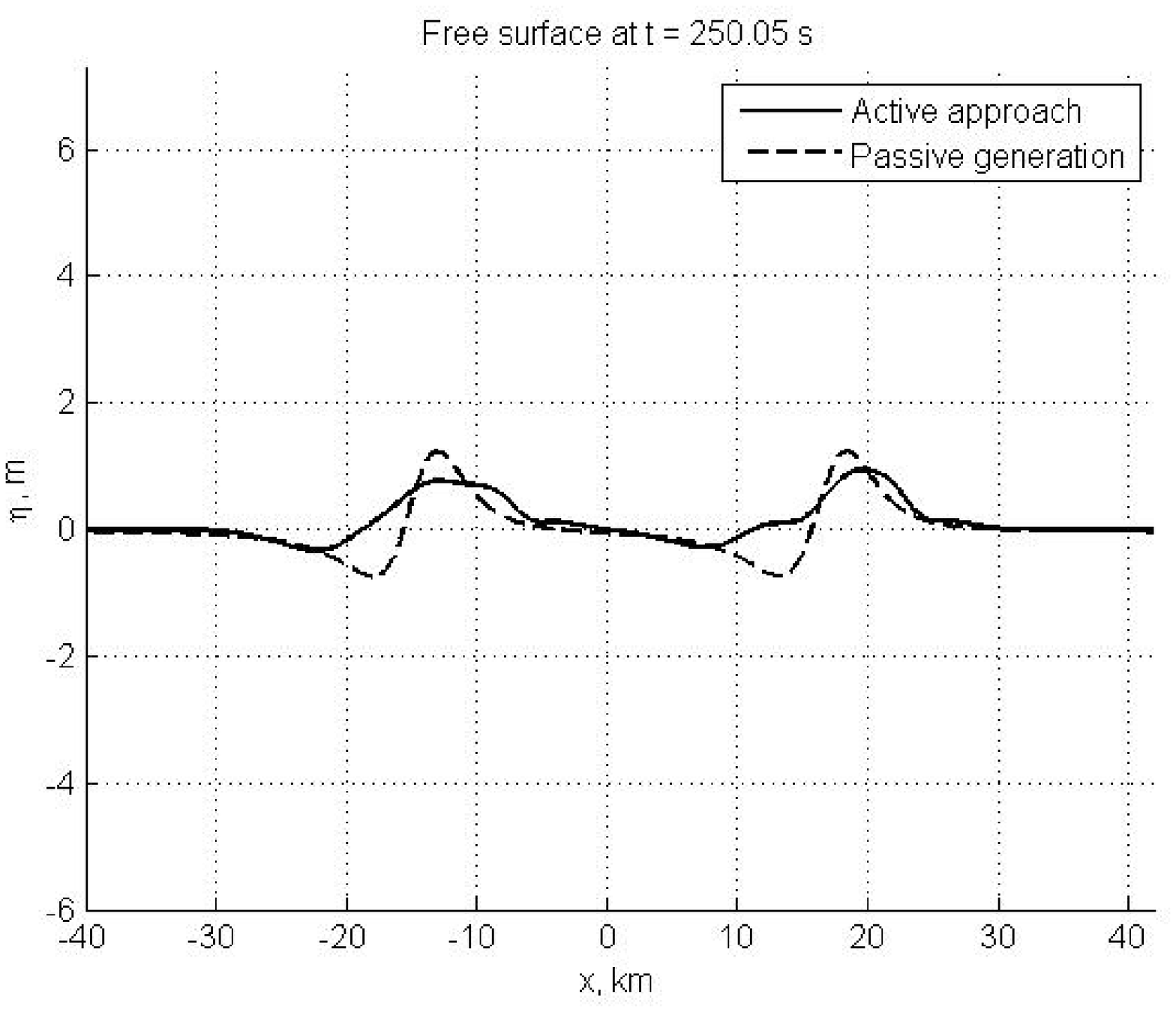, width=7cm} \epsfig{file=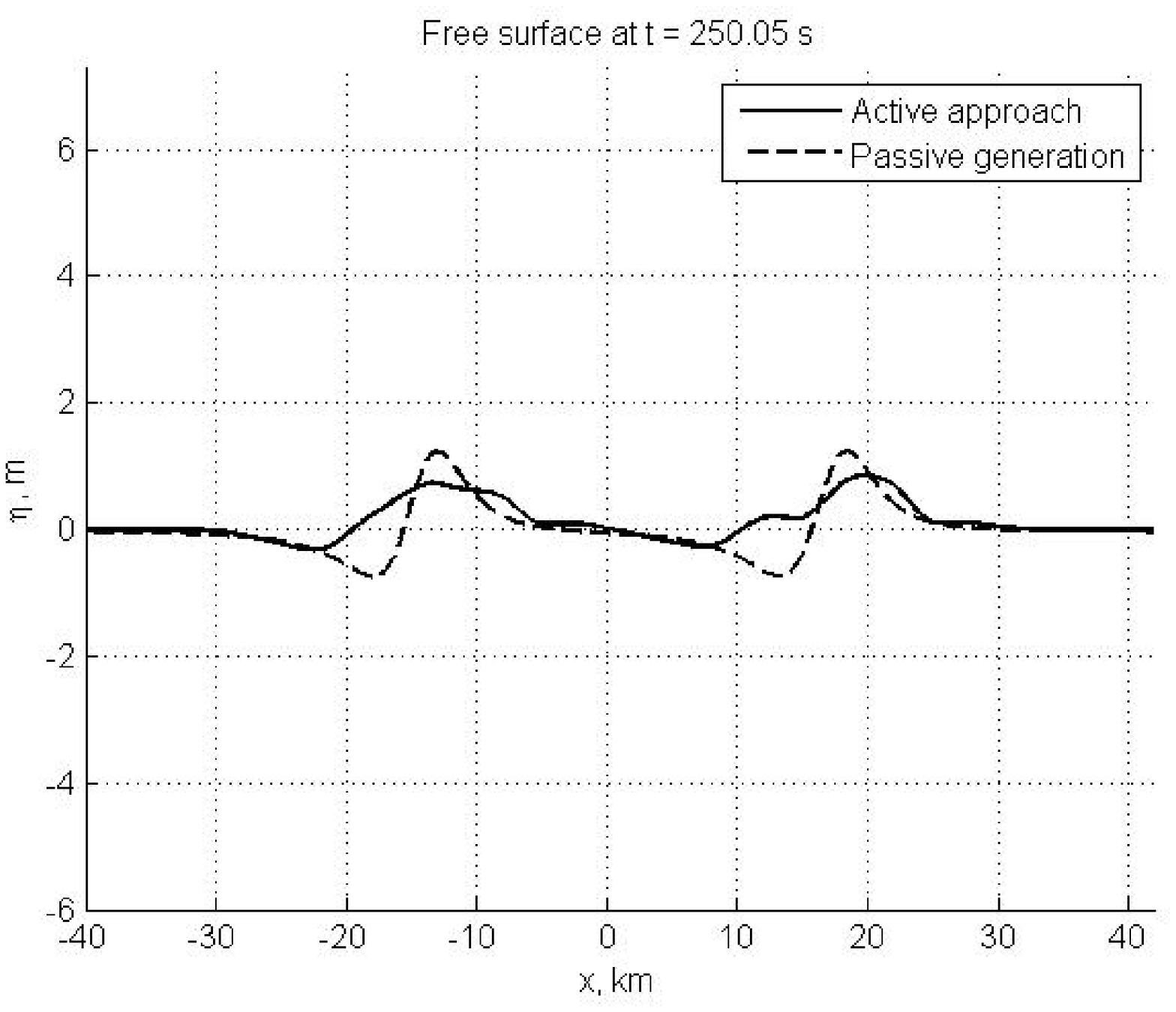, width=7cm}
	\caption{Same as figure \ref{fig:sf150} at time $t=250$s.}
	\label{fig:sf250}
\end{figure}

The second comparison focusses on the influence of the rupture velocity at two separate times
(Figures \ref{fig:sf150} and \ref{fig:sf250}).
The differences between the fast and the relatively slow rupture velocities are small. 

\begin{figure}
	\centering 
 \epsfig{file=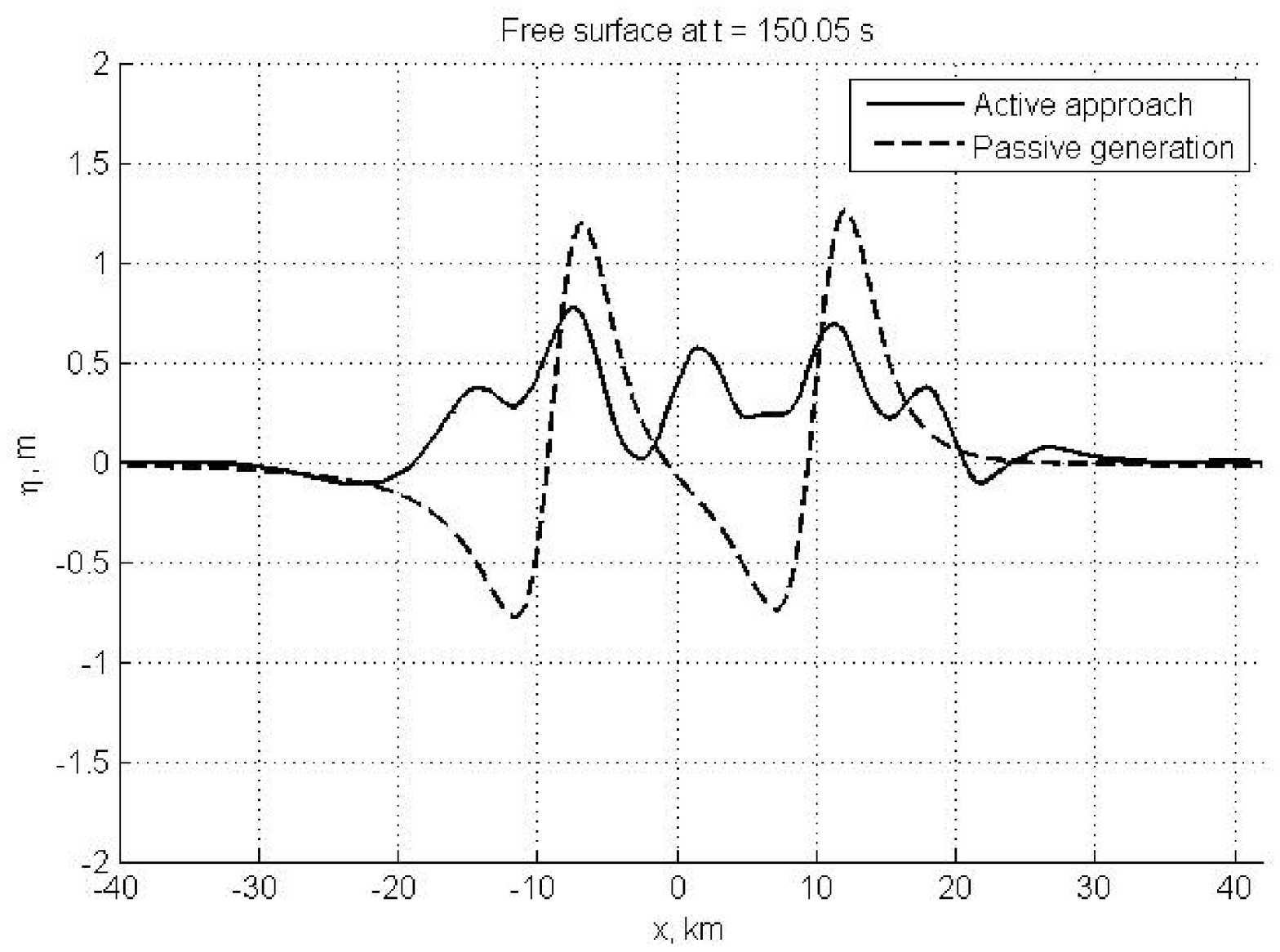, width=7cm} \epsfig{file=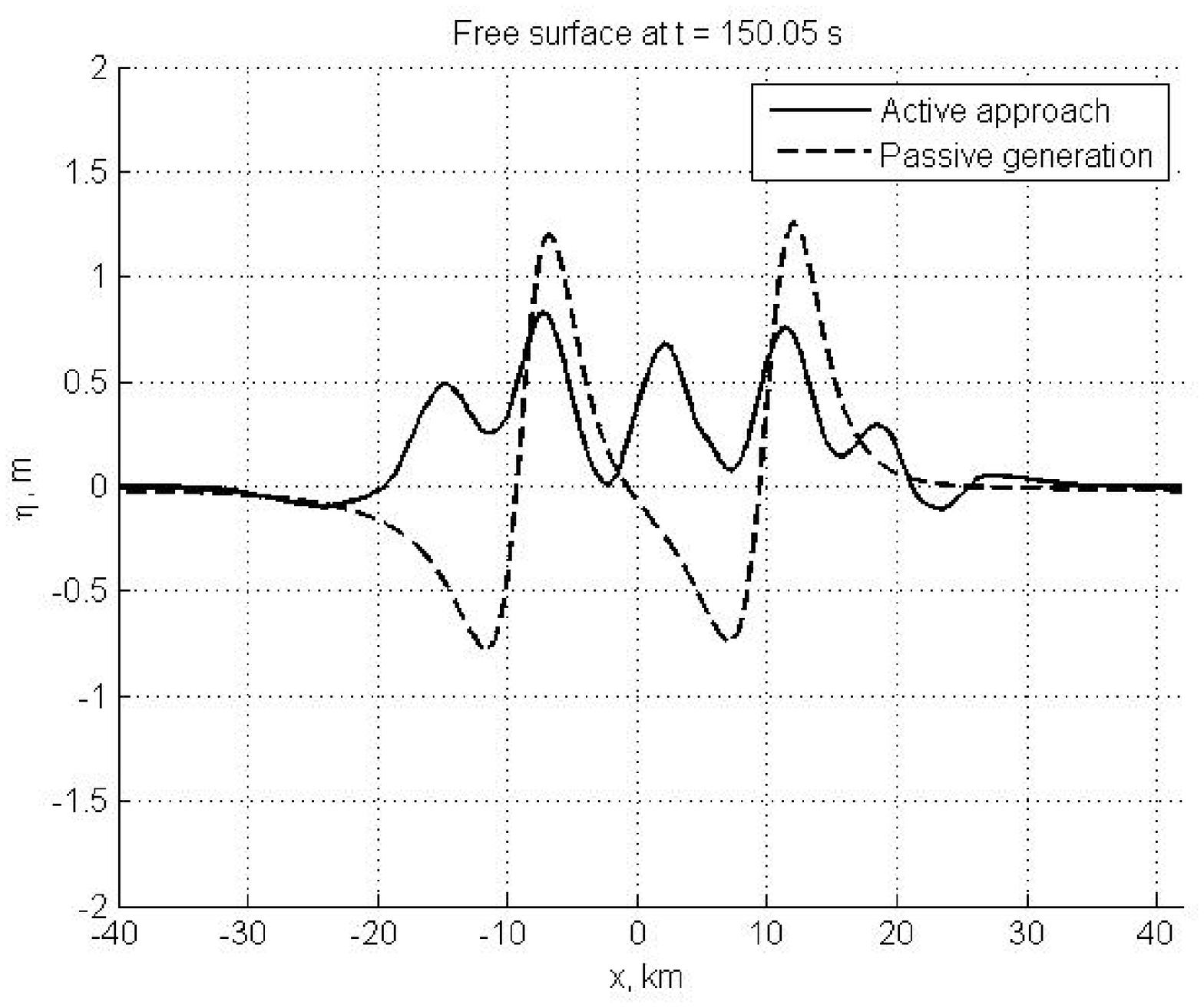, width=7cm}
	\caption{Water free surface at $t=150$s for a longer earthquake. The motion stops after 15s. The left plot compares 
the infinite rupture velocity solution with the passive approach. The right plot compares the case of a slow rupture
($V=1$ km/s) with the passive case.}
	\label{fig:longearth}
\end{figure}

The most interesting comparison is the third one, which focusses on the duration of the earthquake. Recall that
our somewhat artificial definition of earthquake duration is the time at which we stop the bottom motion. After that time, 
the sea bottom remains frozen. 
Figure \ref{fig:longearth} shows the effect of a longer earthquake. One sees that the shapes of the wave train obtained
with the dynamic analysis look more complicated than those obtained with the passive analysis. In particular, the distinction
between leading elevation wave and leading depression wave is not as clear when using the dynamical analysis. It could
be an explanation for the discrepancies between modeled and recorded time series of water levels at various locations
along the California coast for the 1960 Chilean tsunami.

\section{Conclusions}

An approach to model the dynamical character of sea bed deformations during an underwater earthquake was presented.
The governing elastodynamic equations were solved by a finite-element method. The principal
novelty of the present study is the coupling of the resulting displacement field
with the hydrodynamic model.

Two methods for tsunami generation have been compared: static versus dynamic. The computational results
show that the dynamic approach leads to higher water levels in the near-fault area.
These significant differences only occur during
the first instants of the surface deformation and level off later on. However it was also observed that the
shape of the wave train can be altered by dynamical effects. Consequently the distinction between leading elevation
wave and leading depression wave may not be as clear as anticipated. 
Of course the present method is 
computationally more expensive but there is an overall gain in accuracy. Not surprisingly more accurate tsunami computations 
require finer initial conditions such as those obtained by the active generation methodology used in the
present study.

In future work we intend to extend this modeling to three space dimensions since
it is evident that the 2D computations presented in this article have
little interest beyond academics. Moreover we intend to use the exact results of Tadepalli \& Synolakis
\cite{TS1994,TS1996} to check how different the runup may be for the two different initial waves (resulting 
from the passive and dynamic seafloor displacements) in an idealized basin similar to the one used by Okal \&
Synolakis \cite{OS2004}.


\section*{Acknowledgments}
The second author acknowledges the support from the EU project TRANSFER 
(Tsunami Risk ANd Strategies For the European Region) of the 
sixth Framework Programme under contract no. 037058.

\end{document}